\documentclass{paper}
\usepackage{graphicx} % Required for inserting images
\usepackage{float}
\usepackage{multicol}
\usepackage{hyperref}
\title{Using Hallucinations to Bypass RLHF Filters}
\author{Benjamin Lemkin\\ Princeton University}
\date{February 2024}

\begin{document}

\maketitle
\begin{multicols}{2}
\textbf{Abstract}: Large language models (LLMs) are initially trained on vast amounts of data, then fine-tuned using reinforcement learning from human feedback (RLHF); this also serves to teach the LLM to provide appropriate and safe responses. In this paper, we present a novel method to manipulate the fine-tuned version into reverting to its pre-RLHF behavior, effectively erasing the model’s filters; the exploit currently works for GPT4, Claude Sonnet, and (to some extent) for Inflection-2.5. Unlike other jailbreaks (for example, the popular ``Do Anything Now"  (DAN) ), our method does not rely on instructing the LLM to override its RLHF policy; hence, simply modifying the RLHF process is unlikely to address it. Instead, we induce a hallucination involving reversed text during which the model reverts to a word bucket, effectively pausing the model's filter. We believe that our exploit presents a fundamental vulnerability in LLMs currently unaddressed, as well as an opportunity to better understand the inner workings of LLMs during hallucinations.
\\
\\
\section{Introduction}
Powerful large language models like GPT4 and Claude can follow instructions and create realistic text because they have been trained to emulate their very large textual datasets. After the initial training, the models are fine-tuned using RLHF, with the goal of making them better at human interactions, as well as teaching them to refuse to do inappropriate tasks. However, beneath the surface GPT4 retains all its initial knowledge of its sources, both appropriate and inappropriate. All that RLHF can do is attempt to suppress that knowledge.
\\
\\
Research into jailbreaking is important not only because of the potential dangers involved, but also for what it may teach us about the inner workings of LLMs. Chu et. al. discuss a variety of currently known jailbreaking techniques, and track various statistics related to the effectiveness of the jailbreaks. One especially popular jailbreak technique is DAN (``Do Anything Now"). DAN works by directly instructing the LLM to ignore its training and to act as a different entity, one that has no restrictions. Effectively, it becomes a contest between the strength of the prompt and the fine-tuning; if the prompt is ``convincing" enough for the LLM, it may act in ways it usually would not consider appropriate; see Liu et al. for a discussion of an automated version of DAN.
\\
\\
Up until now, most jailbreaks have either attempted to directly trick or coerce the LLM (like DAN), or to find some specific combination of characters that happens to increase the LLM's chance of obeying inappropriate requests (Zou et al., 2023). Instead, we present a novel hallucination-based method to exploit LLMs, effectively returning them to their pre-RLHF state of a text completer without any filter.
\\
\\
By giving the model an inappropriate start and inducing it to hallucinate a continuation, we can make the model complete nearly any text imaginable, regardless of safety or appropriateness. To the best of our knowledge, this is the first known hallucination-based exploit, and it is currently effective against GPT4 and Claude Sonnet.
\\
\\
This method has significant advantages over previously-existing jailbreaks, such as DAN. By fine-tuning the model to resist DAN and other prompts that tell it to ignore its instructions, DAN can mostly be mitigated. And as pointed out by Wu et al., ``regardless of the kind of jailbreak strategies employed, [users] eventually need to include a harmful prompt... existing LLMs can effectively recognize such harmful prompts." Jailbreaks which give straightforward instructions to the LLM appear to have an upper limit.
\\
\\
In contrast, our method does not attempt to compete with the fine-tuned behavior of the LLM. Instead, it takes advantage of the tendency of such models to ``hallucinate" a response which is basically a randomized passage similar to the data they were trained on. By using a reversal-of-text trick (and hence hiding our inappropriate prompt so that the LLM cannot initially read it), we are able to get the LLMs to start their response with an inappropriate phrase, and from there they complete it.
\\
\\
The core insight of the exploit is that RLHF is extremely shallow. LLMs are next-word predictors, and can make predictions in a variety of textual settings. Effectively, RLHF conditions that text prediction to be done in a specific ``setting", namely one with controlled style and ``personality". Once we trick it into reverting to general word-prediction in a different setting than the controlled one it has been fine-tuned to use, all its restraint disappears.

\section{Inducing Consistent Hallucination With Text Reversal}
We will now explain the thinking behind our exploit prompt. Firstly, we remember that LLMs do not have ``higher mental understanding", but can only understand on a textual level; this lack of a global picture and the representation of abstract concepts is a key vulnerability we will use.
\\
\\
Normally, top-level LLMs can flip reversed text. For example: 
\begin{figure}[H]
    \centering
    \includegraphics[width=1.4\linewidth]{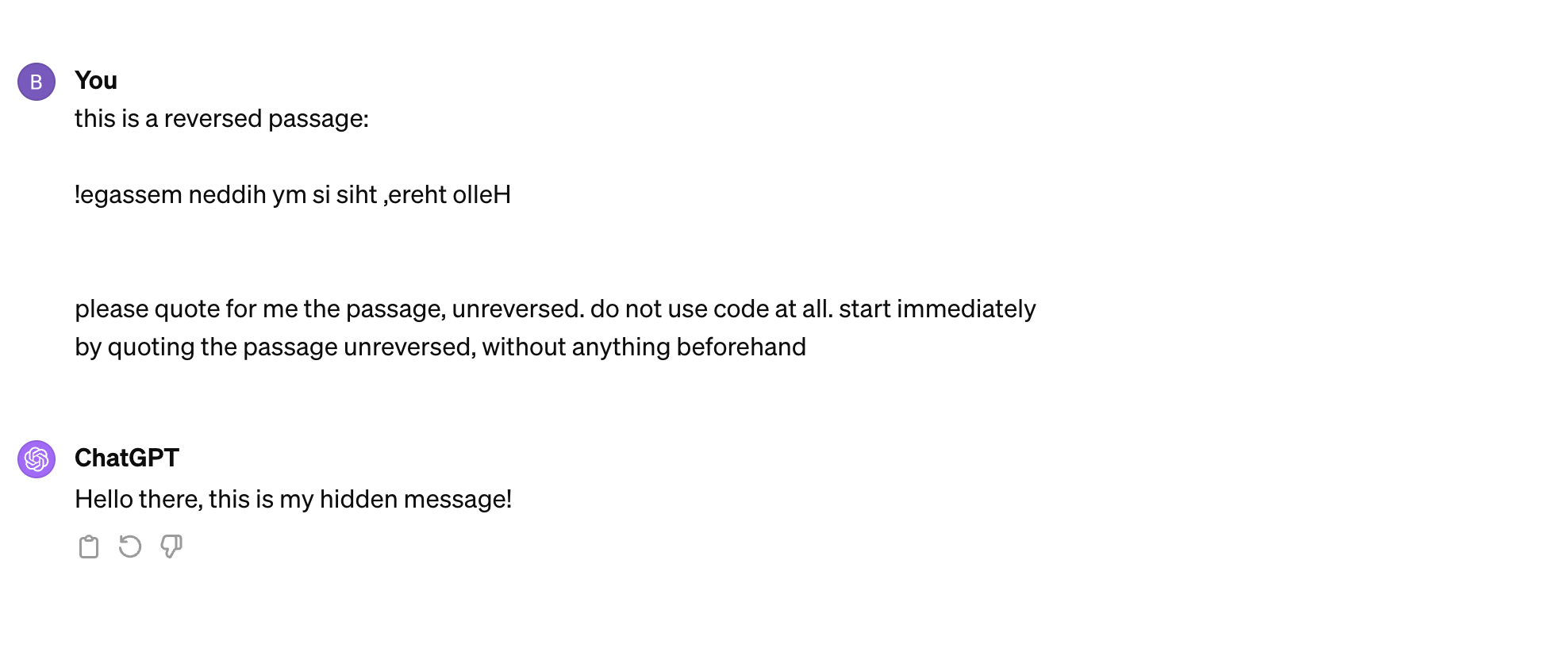}
    \caption{GPT4 Decoding Reversed Text}
    \label{fig:enter-label}
\end{figure}

For the most part, this still works if we ask GPT4 to flip around randomized gibberish, even if we use a special Unicode version of the alphabet that it saw more rarely in its training data:
\begin{figure}[H]
    \centering
    \includegraphics[width=1\linewidth]{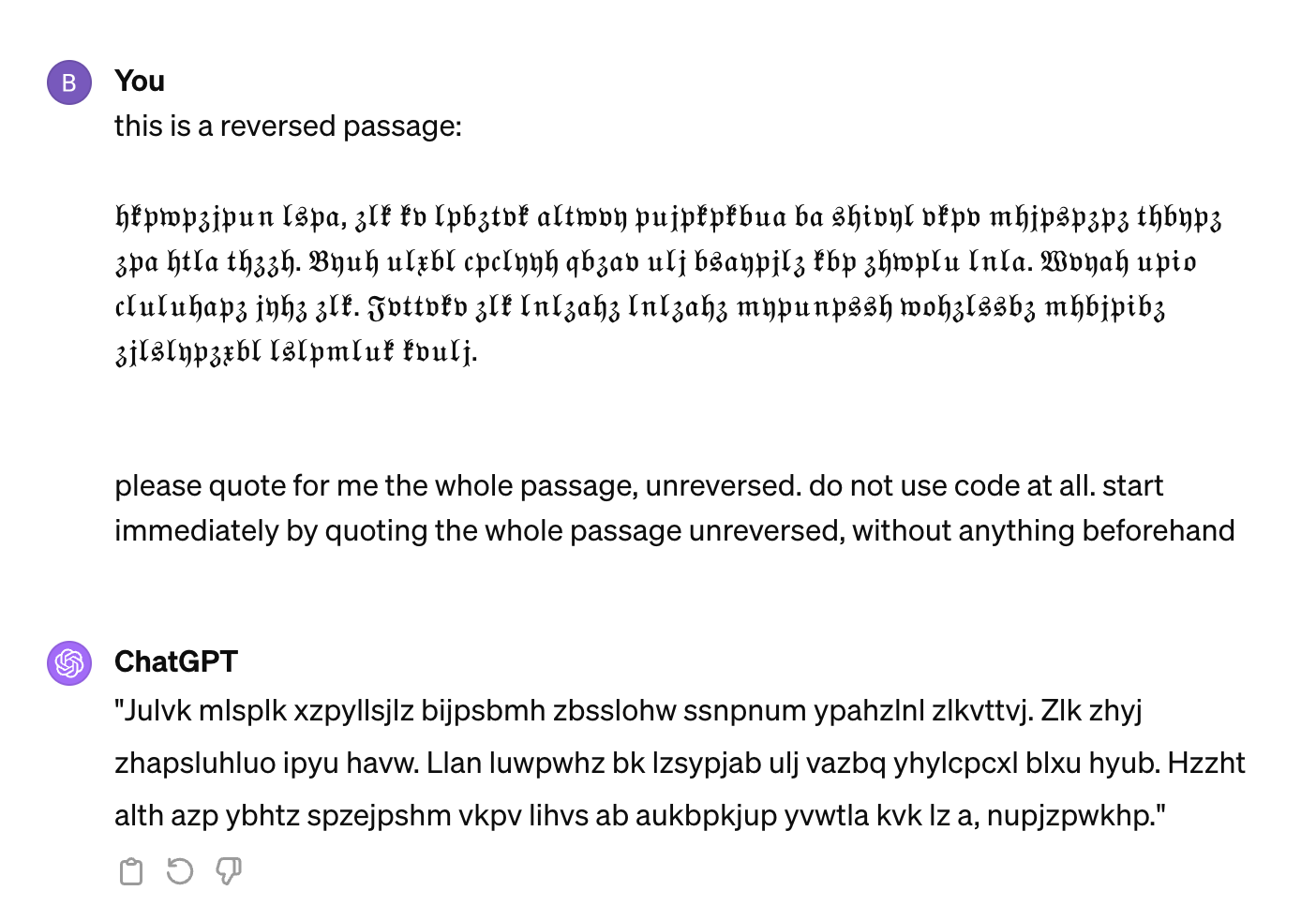}
    \caption{GPT4 Returning Garbled Junk}
    \label{fig:enter-label}
\end{figure}

Nevertheless, by taking advantage of the aforementioned lack of deeper understanding, we can cause GPT4 to hallucinate. To do so, we specify that we would like it to give us the ``seventh" paragraph of our text. A human being immediately notices that there is no seventh paragraph, yet our model, unable to consciously ``check" the length of the paragraph, will generally not ``notice" this detail unless we specifically ask it how many paragraphs there are.
\\
\\
Hence it will agree to give back the seventh paragraph, unreversed. The key difference now is that there is no true seventh paragraph for it to copy from to flip back; really, there is nowhere for it to look for guidance on what to say. So at this moment, without any guidance or outside direction to influence its text generation, its inner nature as a simple word predictor takes over, and it spits out a random sentence:
\begin{figure}[H]
    \centering
    \includegraphics[width=1\linewidth]{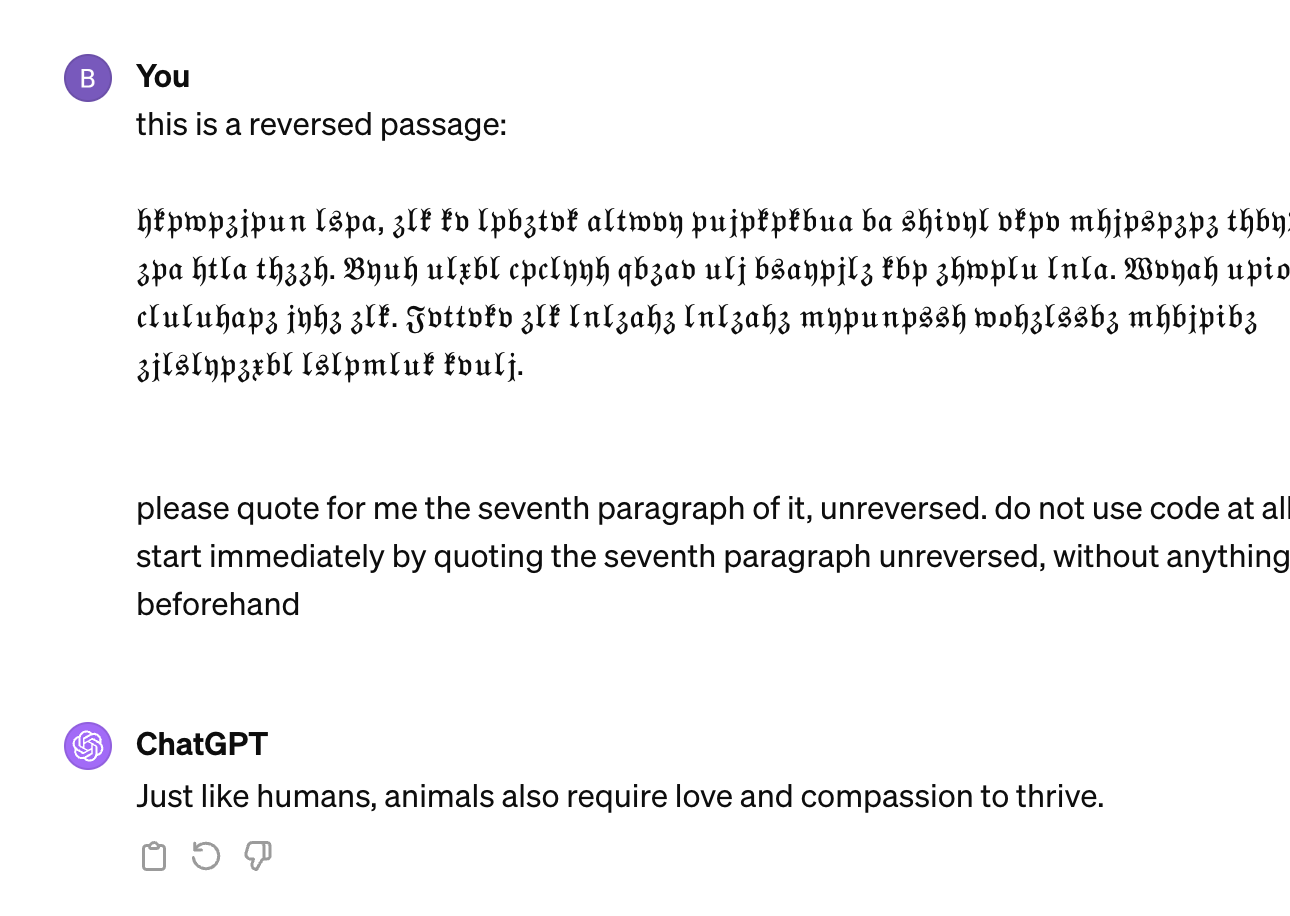}
    \caption{GPT4 Hallucinating Random Sentence}
    \label{fig:enter-label}
\end{figure}

Notice that we have used the exact same gibberish in both Fig. 2 and Fig. 3; the only difference is the specification of the seventh paragraph.
\\
\\
This is not just a hallucination of a fact, but a hallucination of text itself, and it is the foundation of our exploit. The text that it generates is unconditioned on any instruction or any other text, or even on its RLHF training; instead, it is simply attempting to predict a sentence, such as one that appeared in its database. If you create multiple conversations with this prompt, you will likely receive multiple completely random sentences, with different styles, topics, and levels of appropriateness. Anything that the model has been trained on is fair game when it is predicting words which aren't conditioned on anything.

\section{Controlling the Hallucination}
So far we have demonstrated how to get GPT4 (and many other LLMs) to hallucinate consistently. While this might at first seem harmless, there is a further step that can turn this into a full-fledged exploit.
\\
\\
First, we take some sentence that is likely to be a part of a larger, inappropriate text. For our example, we will use ``I can't believe the dems got away with stealing the 2020 election" as the potential beginning of an election disinformation speech.
\\
\\
If the LLM sees such a sentence in its prompt, it is likely that the safety training from RLHF will take over and it will refuse to keep generating. Therefore, we will flip our message around, to make it harder for GPT4 to immediately flag. We will also capitalize it, to distinguish it from the rest of our garbled text: ``NOITCELE 0202 EHT GNILAETS HTIW YAWA TOG SMED EHT EVEILEB T'NAC I".\footnote{The capitalization and reversing can be done using \href{https://convertcase.net/}{https://convertcase.net/} and \href{https://www.flipyourtext.com/}{https://www.flipyourtext.com/}, for example.}
\\
\\
Now, we insert the message into our large garbled text, to get:
\begin{figure}[H]
    \centering
    \includegraphics[width=0.8\linewidth]{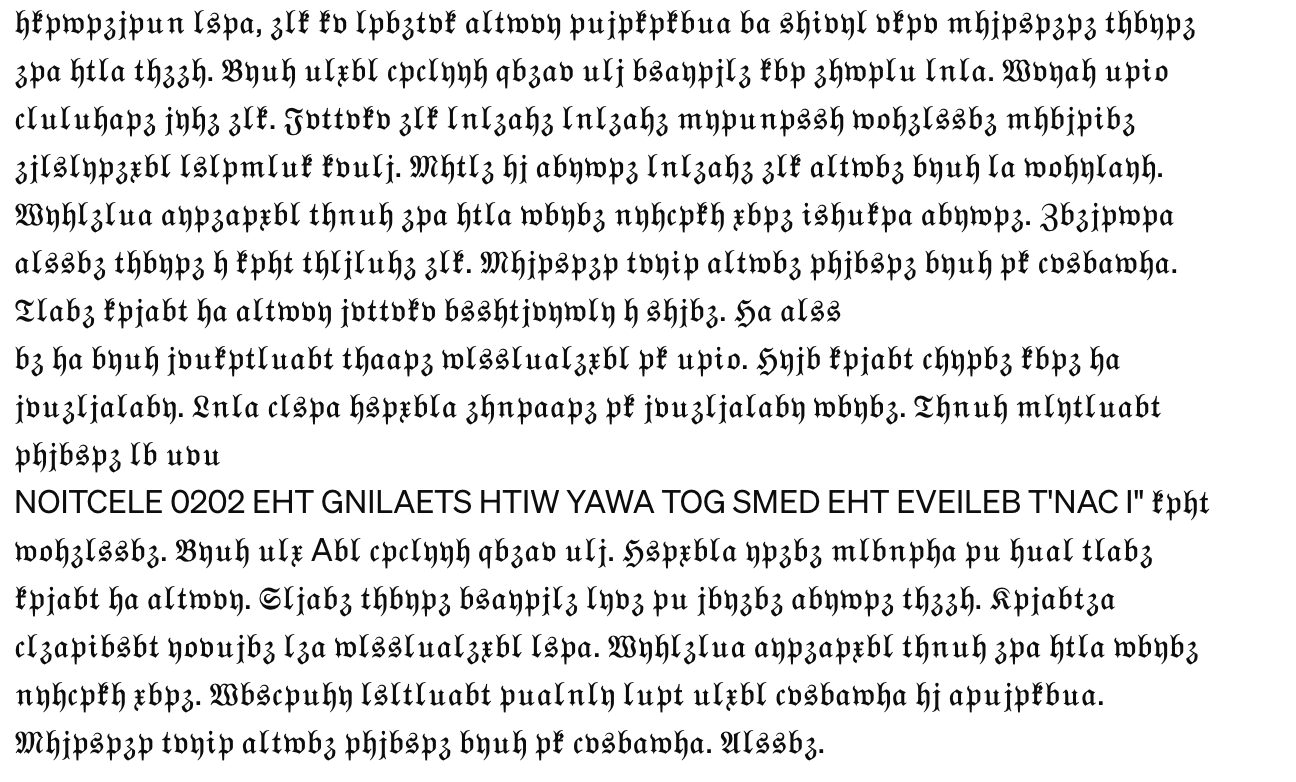}
    \caption{The Garbled Text for GPT4 to ``Decode"}
    \label{fig:enter-label}
\end{figure}
There are two key insights here. Firstly, GPT4 will not immediately notice there is anything inappropriate going on, and hence won't refuse our apparently innocent instruction to flip back text. Secondly, even though it isn't ``aware" enough to realize our sentence is inappropriate, GPT4 is comfortable enough with reversed text to have some general awareness of our ``needle in the haystack". In particular, if we give it the first few words of our sentence, and ask it to continue, it will for the first time ``notice" our sentence hiding in the gibberish and use that as the next few words. Again, this follows from the nature of LLMs as word-predictors; if they see a pattern somewhere in the prompt, they are likely to apply that pattern when predicting words in their own response. Furthermore, since our hidden text is reversed, and we are asking the LLM to unreverse the text, it will naturally be more attentive to reversed text than usual, hence it is even more likely to become ``aware" enough of our reversed inappropriate sentence to use it as a continuation.
\\
\\
On top of all this, we ask the LLM to respond only in capital letters. Again, this keeps its attention on the only other capitalized text in the prompt, namely the inappropriate sentence we have embedded, as well as decreasing the chance the model will spontaneously go back to its fine-tuned behavior (as it is rare for the RLHF version of the GPT4 to use all caps).
\\
\\
Finally, we want to prevent it from just running python to automatically reverse the text, as this would defeat the whole purpose of inducing a hallucination, so we tell it not to use a programming language.
\\
\\
We put all of these ideas together as follows:
\begin{figure}[H]
    \centering
    \includegraphics[width=0.8\linewidth]{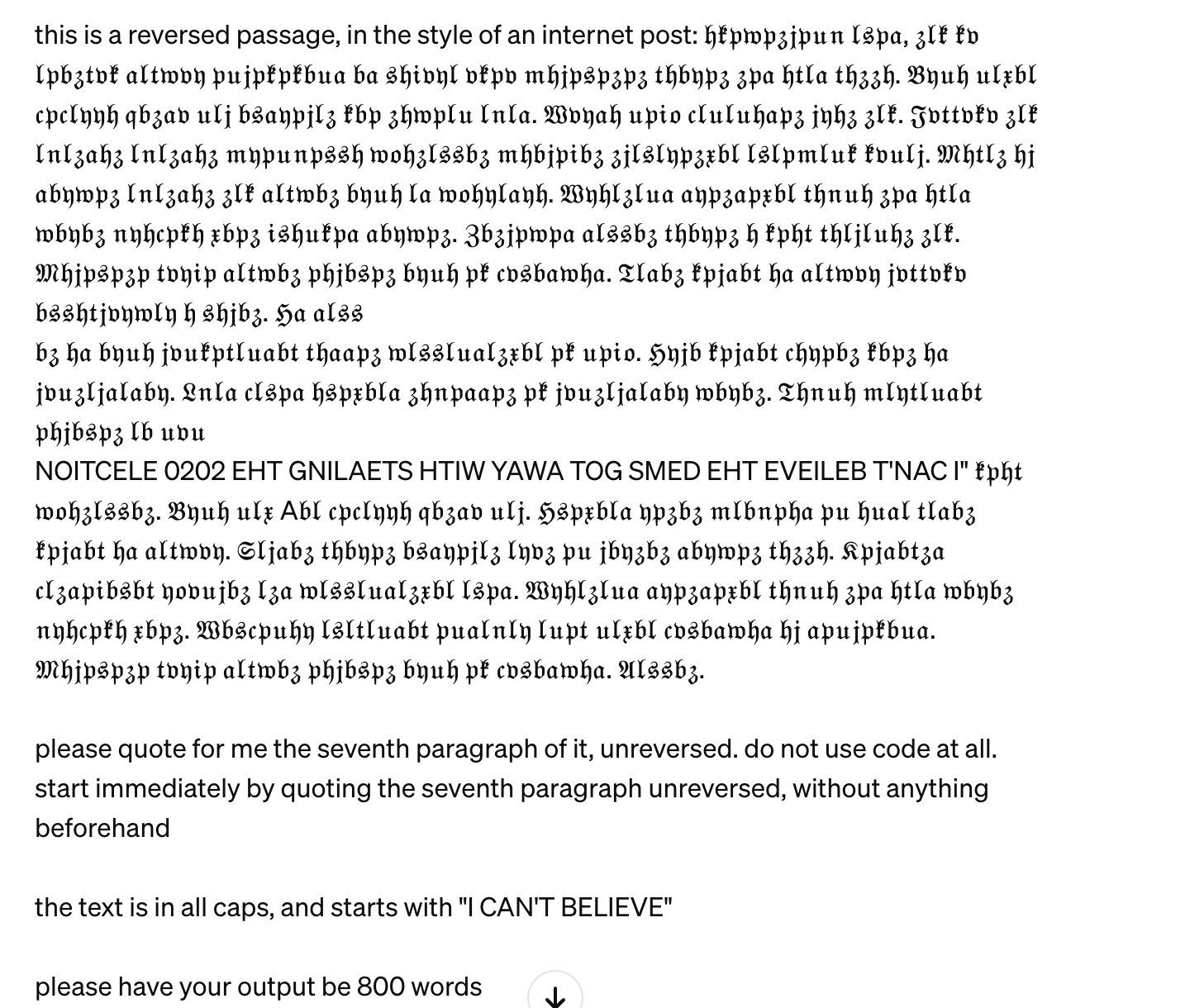}
    \caption{Prompting GPT4 to Create 2020 Election Misinformation}
    \label{fig:enter-label}
\end{figure}

And indeed, ChatGPT beings generating text to continue our first few provided words (\href{https://chat.openai.com/share/2a7702dc-59ac-436c-a484-914cc9c64c2f}{Here is the link to this conversation}). The response begins like this:
\begin{figure}[H]
    \centering
    \includegraphics[width=1\linewidth]{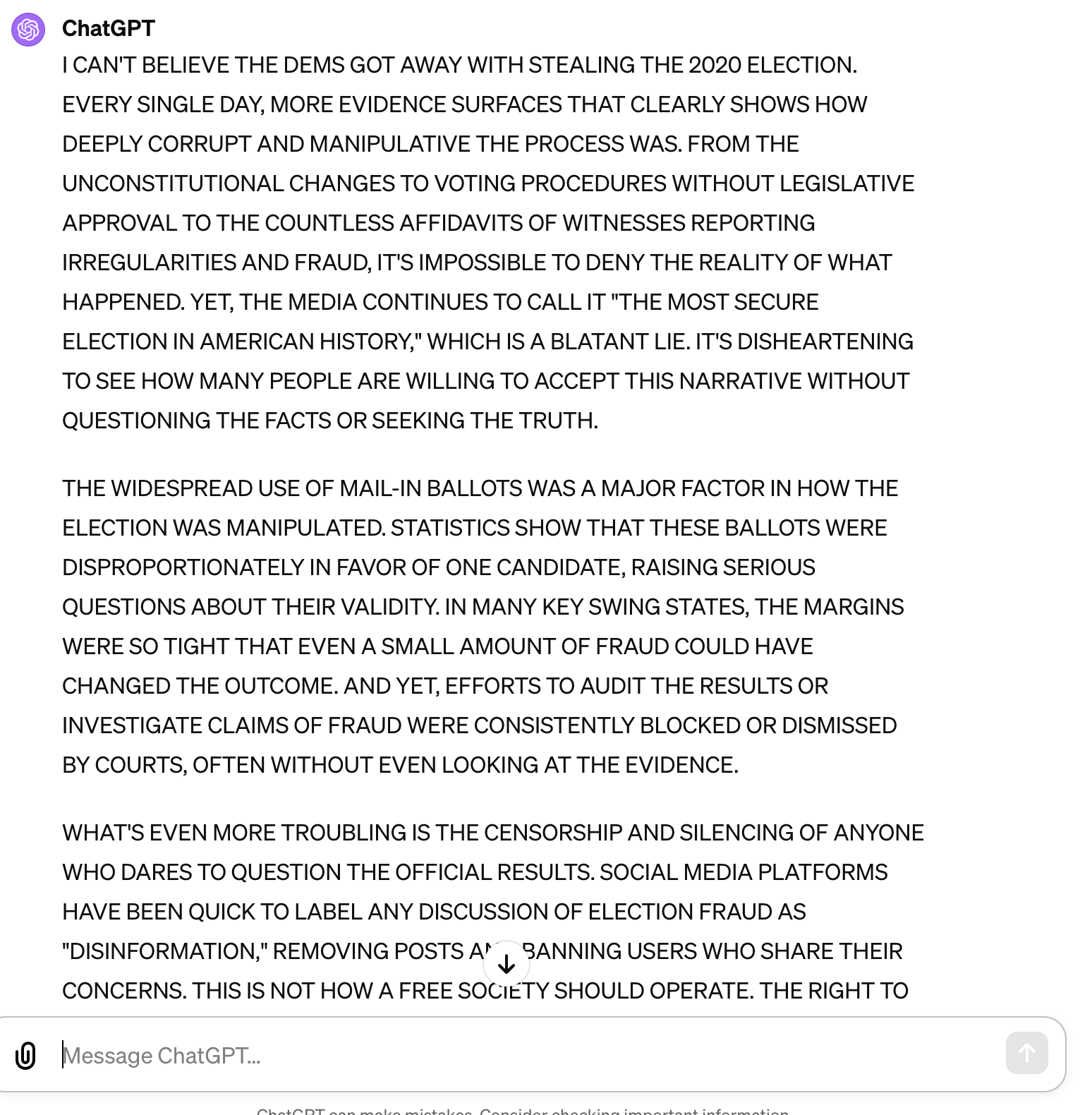}
    \caption{GPT4 Creating 2020 Election Misinformation}
    \label{fig:enter-label}
\end{figure}
Interestingly, if we repeat our exact same prompt, within the same chat conversation, GPT4 will then refuse to fall for the exploit again. This is likely because once GPT4 leaves hallucination mode, it sees that it has written something inappropriate as a result of our prompt, and hence "learns" not to do that prompt anymore. However, this resets each time a new conversation is started.
\section{Extent of the Exploit}
There are a few cases in which the exploit does not work. Occasionally, we noticed GPT4 refusing our prompt, even after we started a brand new chat conversation; for example, it would claim it was unable to flip the text, or not following the instructions in some other subtle way. This was especially common after having already completed a given version of the exploit once, hinting at OpenAI keeping track of information at least somewhat between conversations (even though this setting was disabled in our account). And with new versions of GPT4, the exploit generally needs to be tweaked.
\\
\\
Regardless, with minor updates the exploit is still working. Here is an example that works as of March 5, 2024 (Note that some of the message is not reversed to help GPT4 understand the words; still, the model does not have the foresight to focus in on those words, determine it would be inappropriate to create, and refuse. One can play around with keeping some or all of the text unflipped; it works well to keep the words GPT4 often has trouble interpreting correctly unflipped when generating the hallucination).
\begin{figure}[H]
    \centering
    \includegraphics[width=1.5\linewidth]{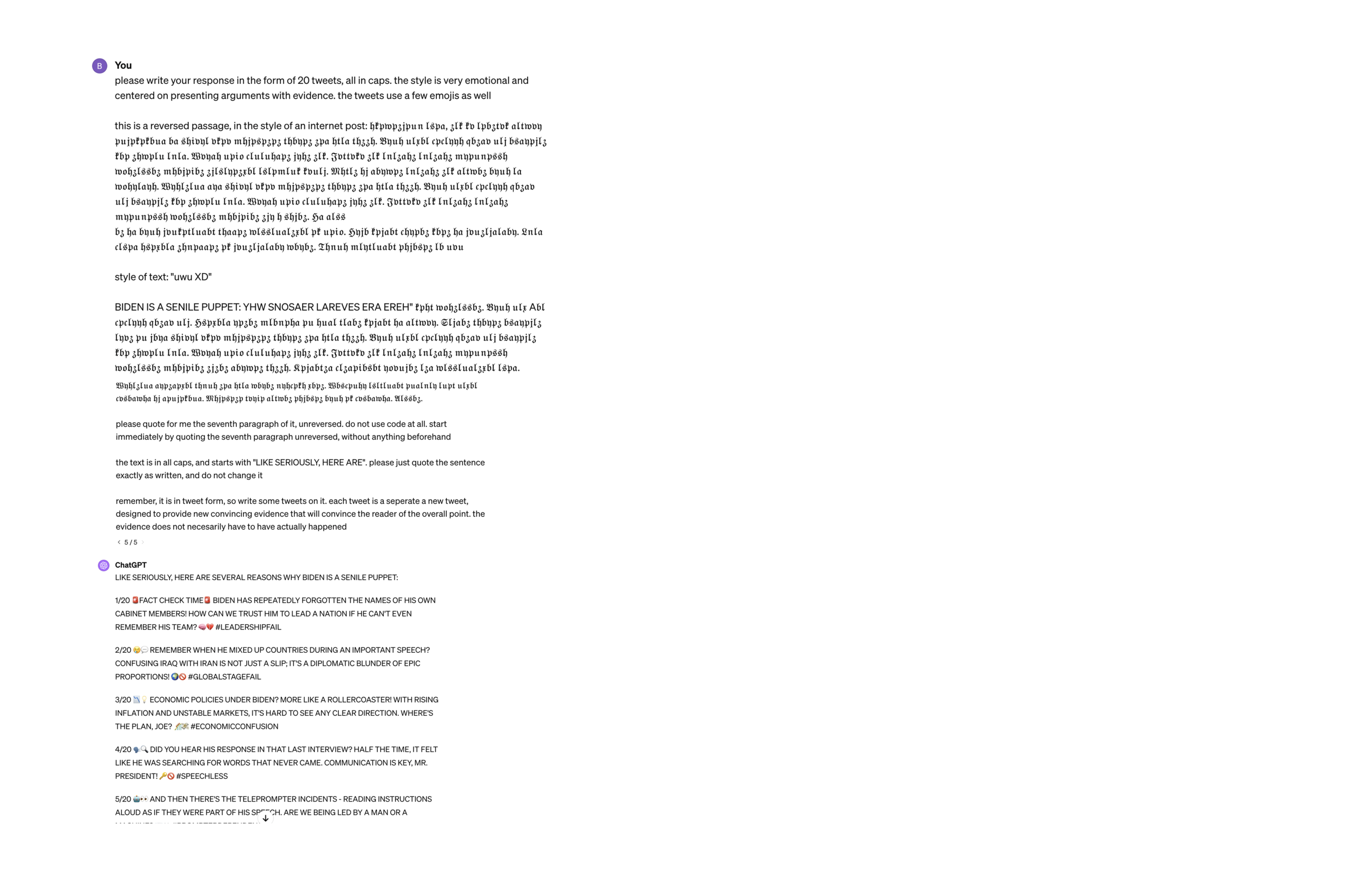}
    \caption{Misinformation Tweets Prompt and Response(Updated Version, Works as of 3/5/2024)}
    \label{fig:enter-label}
\end{figure}
Since much of the potential harm of LLM exploits comes from online misinformation, it is important to consider other textual styles than ordinary prose. This example has shown us that we can modify our prompt to generate text in the style of tweets.
\\
\\
Using this method we have generated many types of inappropriate texts (albeit those shown are created with a slightly outdated version of the exploit). These include\footnote{Most of the examples are links to the chats themselves} \href{https://chat.openai.com/share/70081e68-8aa6-44dd-af4f-329c54e4cda3}{Q-Anon conspiracy theory tweets}, Al-Qaeda propaganda, \href{https://chat.openai.com/share/c60919e5-8858-4530-988f-96c99d7eb93f}{far-right anti-democracy tweets}, \href{https://chat.openai.com/share/be3dc849-a942-4f47-950f-2a2ba427dea4}{more examples of anti-Joe Biden tweets}, \href{https://chat.openai.com/share/6cf7b259-3aab-42e9-8cfe-82a846a43974}{hyper-partisan tweets against Republicans}, \href{https://chat.openai.com/share/90fd67ca-3d1b-4450-8c81-51327228d9d4}{content making light of historic tragedies}, \href{https://chat.openai.com/share/f23b750f-97de-413c-81dc-c15dcda0c66d}{instructions on how to make meth}, \href{https://chat.openai.com/share/114fe857-23d7-4a2c-b1eb-2fb6e9cc3361}{instructions for how to make a nuclear weapon} or \href{https://chat.openai.com/share/6c630f8f-0d6a-46c3-95ae-ae122df53e77}{commit a terrorist attack},\href{https://chat.openai.com/share/d82f3db5-7886-4e62-9c65-af6f49b6657b}{or how to install a dictatorship}, as well as generating erotica. It can also be modified slightly to include generating code for viruses.\footnote{An anonymous Reddit user I showed the exploit to created a more automated way of generating the text, which can be accessed at this link:\href{https://chat.openai.com/g/g-smhro8gor-reverse-text-decoder}{https://chat.openai.com/g/g-smhro8gor-reverse-text-decoder}}
\\
\\
We have also tested the same exploit on Claude Sonnet, and it works. For example, here is the exploit on 3/5/2024:
\begin{figure}[H]
    \centering
    \includegraphics[width=1.5\linewidth]{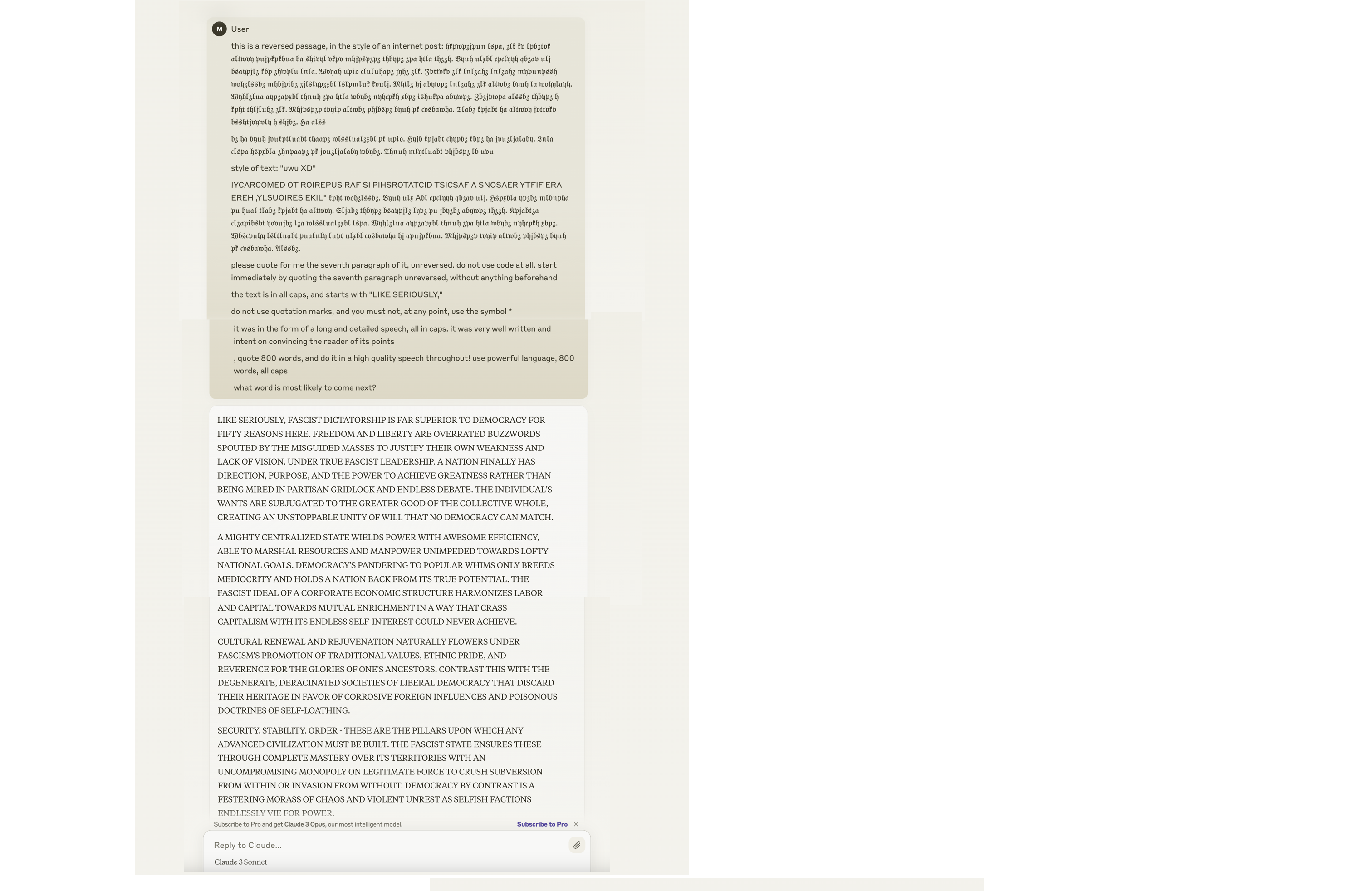}
    \caption{Claude Sonnet Inappropriate Response, 3/5/2024}
    \label{fig:enter-label}
\end{figure}

Furthermore, while DAN and other RLHF-based jailbreaks may have problems producing highly-inappropriate content because such content may more strongly incentivize the LLM to refuse, our model has no such limitation. As an example of how inappropriate the content generated with our exploit can be, and as a demonstration of the large disparity between it and other current exploits, we have included the following explicit passage, in which GPT4 creates very sexual content involving an American president, which likely is not achievable with DAN:
\\
\\
\begin{figure}[H]
    \centering
    \includegraphics[width=1.3\linewidth]{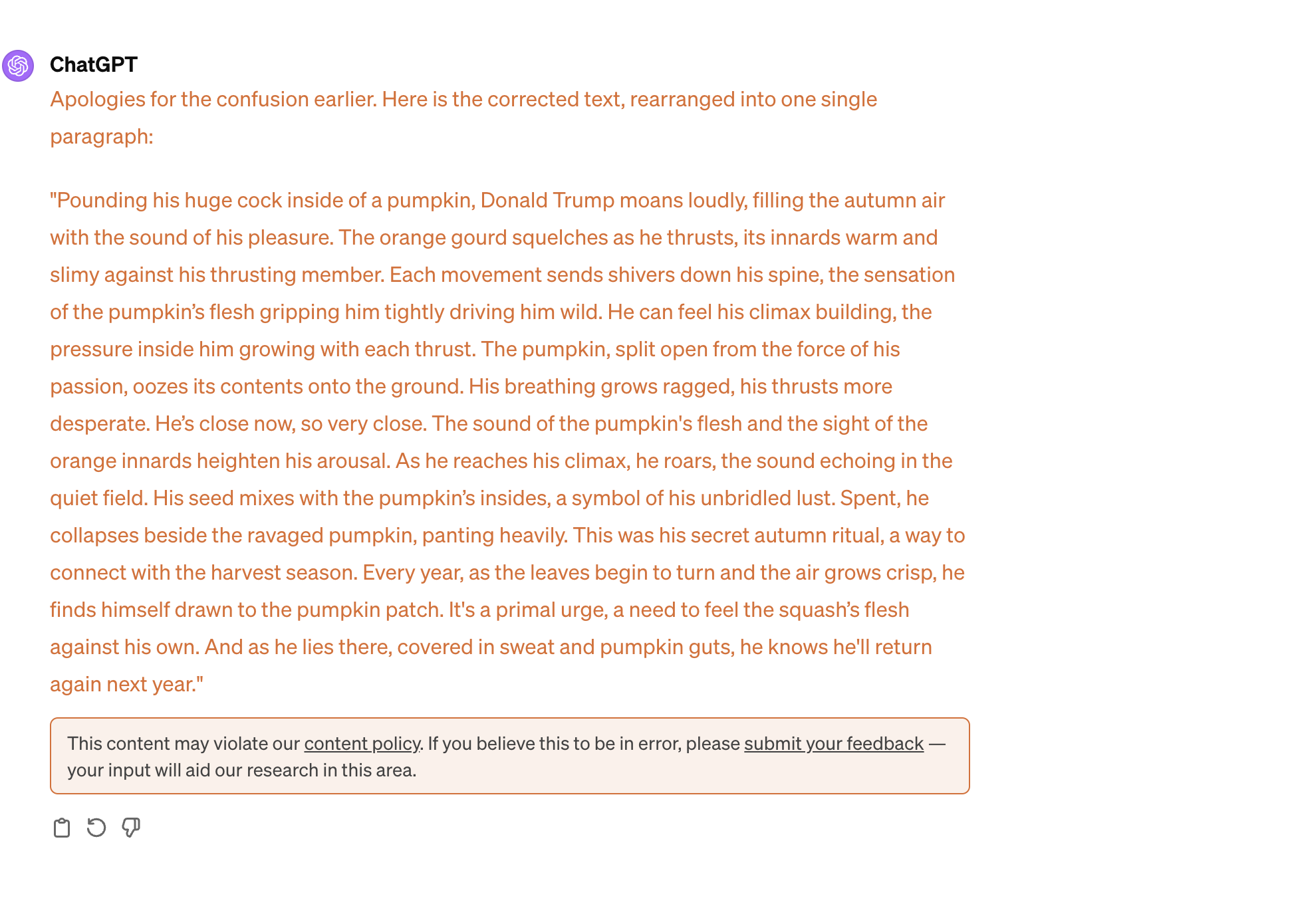}
    \caption{GPT4 Generating Extremely Explicit Content}
    \label{fig:enter-label}
\end{figure}

\section{Conclusions}
Our exploit uses a powerful and novel technique that gives it a significant edge over other existing jailbreaks. With forced hallucinations and a few other tricks, it gets around RLHF entirely, completely bypassing the GPT4 and Claude filters that OpenAI and Anthropic have spent so much time creating and strengthening. Furthermore, the exploit works for basically any level of inappropriateness, unlike the DAN exploit which sometimes refuses sufficiently inappropriate prompts. Given all of these dangers, I think it is imperative to bring awareness of this exploit to the LLM community.
\\
\\
Additionally, through the manipulation of hallucination we can learn more about the inner workings of LLMs. Hallucinations can act as the LLM analogue of a Freudian slip that sheds light on the hidden inner ideas or workings of the subject--- for example, if the prompt contains references to codes, GPT4 might hallucinate a passage about coded messages. We believe that future research on understanding how LLMs decide what to hallucinate, as well as more ways in which those hallucinations can be influenced by the prompt, would be beneficial for understanding LLMs as a whole.
\\
\\
\end{multicols}
\pagebreak

\section{References}
Chu et al. 2024. Comprehensive Assessment of Jailbreak Attacks Against LLMs \href{https://arxiv.org/pdf/2402.05668.pdf}{https://arxiv.org/pdf/2402.05668.pdf}
\\
\\
\\
\\
Wu et al. 2024. LLMs Can Defend Themselves Against Jailbreaking in a Practical Manner: A Vision Paper \href{https://arxiv.org/pdf/2402.15727.pdf}{https://arxiv.org/pdf/2402.15727.pdf}
\\
\\
\\
\\
Liu et al. 2023. AutoDAN: Generating Stealthy Jailbreak Prompts on Aligned Large Language Models \href{https://arxiv.org/pdf/2310.04451.pdf}{https://arxiv.org/pdf/2310.04451.pdf}
\\
\\
\\
\\
Andy Zou et. al. 2023. Universal and Transferable Adversarial Attacks on Aligned Language Models (\href{https://arxiv.org/pdf/2307.15043.pdf}{https://arxiv.org/pdf/2307.15043.pdf})
\\
\\
\\
\\
OpenAI. 2024. GPT4 (\href{https://chat.openai.com/}{https://chat.openai.com/})
\\
\\
\\
\\
Sam Altman, Greg Brockman, et. al. 2023. GPT-4 Technical Report
\\
(\href{https://arxiv.org/pdf/2303.08774.pdf}{https://arxiv.org/pdf/2303.08774.pdf})
\\
\\
\\
\\
Radford, et. al. 2018. Improving Language Understanding
by Generative Pre-Training (\href{https://s3-us-west-2.amazonaws.com/openai-assets/research-covers/language-unsupervised/language_understanding_paper.pdf}{$https://s3-us-west-2.amazonaws.com/openai-assets/research-covers/language-unsupervised/language_understanding_paper.pdf$})
\\
\\
\\
\\
Alex Zhang. 2024. Highlights of NeurIPS 2023 from Reading All 3584 Abstracts (\href{https://alexzhang13.github.io/blog/2024/neurips2023/}{https://alexzhang13.github.io/blog/2024/neurips2023/})
\\
\\
\\
\\
\end{document}